\documentclass[%
 ,twocolumn%
 ,secnumarabic%
,amssymb, amsmath,nobibnotes,prl, aps]{revtex4}

\newcommand{\ak}{{\mid\!\!\bk\!\!\mid}}
\newcommand{\bk}{{\bf k}}
\newcommand{\bb}{{\bf b}}

\newcommand{\eps}{{\epsilon}}

\newcommand{\psib}{{\bar{\psi}}}

\newcommand{\omegan}{\omega_{n}}

\expandafter\ifx\csname package@font\endcsname\relax\else
 \expandafter\expandafter
 \expandafter\usepackage
 \expandafter\expandafter
 \expandafter{\csname package@font\endcsname}%
\fi

\begin{document} \draft

\title{Quantum criticality of $d$-wave quasiparticles and 
superconducting phase fluctuations}%
\author{Oskar Vafek and Zlatko Te\v sanovi\'c}
\address{Department of Physics and Astronomy, Johns Hopkins
University, Baltimore, MD 21218, USA 
\\ {\rm(\today)}
}
\begin{abstract}
\medskip
We present finite temperature ($T$) extension of 
the QED$_3$ theory of underdoped cuprates.
The theory describes nodal quasiparticles 
whose interactions with quantum proliferated $hc/2e$ vortex-antivortex pairs
are represented by an emergent U(1) gauge field.
Finite $T$ introduces a scale beyond which
the spatial fluctuations of 
vorticity are suppressed. As a result, the  
spin susceptibility of the pseudogap state
is bounded by $T^2$ at low $T$
and crosses over to $\sim T$ at higher $T$, while 
the low-$T$ specific heat scales as $T^2$, reflecting
the thermodynamics of QED$_3$. The Wilson ratio vanishes as $T\to 0$;
the pseudogap state is a ``thermal (semi-)metal'' but a ``spin/charge
dielectric.''
This non-Fermi liquid behavior originates from two general principles:
spin correlations induced by 
``gauge'' interactions of quasiparticles
and fluctuating vortices and the ``relativistic'' scaling 
of the $T=0$ fixed point. 
\end{abstract}

\maketitle
Recent experiments \cite{corson,ong,campuzano}
provide support for the view that
the pseudogap state of copper oxides
represents a phase-disordered superconductor \cite{emerykivelson}.
This view is central to the QED$_3$ 
theory of underdoped cuprates \cite{ftqed}
whose degrees of freedom, Bogoliubov-deGennes (BdG)
quasiparticles and
fluctuating $hc/2e$ vortex-antivortex pairs, and
their mutual interactions are
argued to capture the effective low 
energy physics of a 
$d$-wave superconductor in a doped Mott insulator.
Within the pseudogap phase of high-$T_c$ cuprates
this QED$_3$ theory assumes the role
played by the Fermi liquid theory in 
conventional metals and superconductors. The theory
possesses three major dynamical symmetries: relativistic
and gauge invariance and chiral symmetry, all three {\em emergent}
in nature \cite{footiii}. 
Irrespective of whether the {\em symmetric} phase 
of QED$_3$ is the true $T=0$ ground state of the system, 
these symmetries conspire to make its peculiar brand of
quantum criticality a strongly attractive basin of influence 
on physical properties of the pseudogap state.

In this Letter we focus on the subset of such properties which
are intrinsically important 
features of a condensed matter system: thermodynamics 
and spin response of the pseudogap state. Our primary aim is to 
stimulate
experimental activity by providing explicit and testable theoretical
predictions stemming from the theory of Ref. \cite{ftqed}. 
The summary of our
results is as follows: first, in order to derive thermodynamics
and spin susceptibility from QED$_3$ theory
we generalize its form to finite $T$. This is a matter of some 
subtlety
since, the moment $T\not =0$, the theory loses its fictitious 
``relativistic
invariance''. Second, we show that the 
theory predicts a finite
$T$ scaling form for thermodynamic quantities in the pseudogap
state and propose that this form be tested experimentally. Third, we
determine the leading $T\not =0$ scaling and 
demonstrate that the deviations from 
``relativistic invariance'' are actually {\em irrelevant} for
$T$ much less than the pseudogap temperature $T^*$, in the sense
that the leading order $T\not = 0$ scaling of thermodynamic
functions remains that of the {\em finite-$T$
symmetric} QED$_3$ \cite{plasmon}. 
These deviations from ``relativistic invariance''
do, however, affect higher order terms. 
Finally, we evaluate the uniform magnetic spin susceptibility $\chi$ 
and 
show that it is bounded by $T^2$ at low $T$ 
but crosses over to $\sim T$ at higher $T$, closer to $T^*$. 
Consequently, the
Wilson ratio $R=\chi T/c_v$ vanishes as $T\to 0$:
{\em the QED$_3$ theory implies the non-Fermi liquid 
nature 
of the pseudogap state in cuprates}.
Such a state is a thermal (semi-)metal but a
spin and charge dielectric.\cite{footiv}
Our results are thus suggestive of the breakdown of Wiedemann-Franz law
in the pseudogap state.

The spin susceptibility of a 
$d$-wave superconductor vanishes linearly with temperature. 
A way to understand this result is 
to notice that the spin part of 
the ground state wavefunction, being a spin singlet, 
remains unperturbed by the application of a weak 
{\em uniform} magnetic field. However, the 
excited quasiparticle states are {\em not} in general 
spin singlets and therefore contribute to 
the finite temperature susceptibility. 
Because their density of states is linear 
at low energies, at finite temperature
the number of  quasiparticles that 
are excited is $\sim k_B T$, each contributing 
a constant to the 
Pauli-like uniform spin 
susceptibility $\chi$. Thus $\chi \sim T$. 

When the superconducting phase order is destroyed 
by proliferation of unbound quantum vortex-antivortex pairs, 
the low-energy quasiparticles 
are strongly interacting. 
The interaction originates from the fact that
it is the spin singlet {\em pairs} that acquire 
{\em one unit} of angular momentum in their center 
of the mass coordinate, carried by an $hc/2e$ vortex. 
This translates into topological frustration 
in the propagation of BdG ``spinon'' excitations. As a result, 
non-trivial spin correlations persist in the excited 
states of the phase-disordered $d$-wave superconductor. 
At low temperature these correlations can be described
by an emergent U(1) gauge field \cite{ftqed} and lead to suppression 
of $\chi$ relative to its value for non-interacting BdG quasiparticles.
We shall argue below that $\chi \alt T^2$.

Similarly, in a $d$-wave superconductor, 
linear density of the quasiparticle states 
translates into a $T^2$ dependence of the low-$T$ specific heat. 
When the interactions between 
quasiparticles are included the spectral weight is transferred 
to multi-particle states. 
Within QED$_3$ theory, however, the strongly interacting 
IR (infra-red) fixed point possesses emergent ``relativistic'' 
invariance and the dynamical critical exponent $z=1$. 
Furthermore, the effective quantum action for vortices, deep
in the phase-disordered pseudogap state,
introduces an additional lengthscale, the superconducting 
correlation length $\xi_{\tau,\perp}$ (labels $\tau$ and $\perp$
stand for time- and space-like, respectively). At $T=0$ this 
scale serves as a short distance 
cutoff of the theory and is generically doping ($x$) dependent.
We then argue 
that under rather general circumstances
the low-$T$ ($T\ll T^*$) electronic specific 
heat scales as $T^2$ while the free
energy goes as $T^3$. We now proceed to substantiate the above claims.

{\em Vorticity fluctuations:}\\ 
Deep in the phase disordered pseudogap
state the fluctuations in the vorticity 3-vector 
$b_{\mu}=\eps_{\mu\nu\lambda}\partial_{\nu}a_{\lambda}$ 
are described by the following ``bare" Lagrangian:
\begin{multline}
{\mathcal L}_0[a_{\mu}]=\frac{K}{2} f_{\perp}
\left(\frac{T}{\bk},\frac{T}{\omegan},KT,K_{\tau}T \right)\;b_0^2 \\
+ \frac{K_{\tau}}{2} f_{\tau}\left(\frac{T}{\bk},\frac{T}{\omegan},
KT,K_{\tau}T\right)\;\bb^2,
\label{L0}
\end{multline}
where $\omega_n$ is Matsubara frequency,
$K$,$K_{\tau}$ are related to the {\em finite}
superconducting correlation length of the pseudogap state as
$K_{\tau} \propto \xi^2_{sc}/\xi_{\tau}$,
$K \propto \xi_{\tau}$ \cite{ftqed}
and we have also
set $v_F=v_\Delta =1$ for simplicity. 
$f_{\tau,\perp}$ are general scaling functions 
describing how ${\cal L}_0[a_{\mu}]$ is modified from its 
``relativistically
invariant'' $T=0$ form as the temperature is turned on. They
involve only the thermal length $\sim v_F/T$ and $K$,$K_{\tau}$ and 
satisfy
the condition $f_{\tau}(0,0,0,0)=f_{\perp}(0,0,0,0)=1$.
Physically, the modifications embodied in
$f_{\tau,\perp}$ are due to changes in the pattern of
vortex-antivortex fluctuations induced by finite $T$. The explicit
expressions for $f_{\tau,\perp}$ depend on the details of a particular
model for phase disorder within the pseudogap state -- as we emphasize
below, however, our results are either completely insensitive to
such details or reflect only the most general features of $f_{\tau,\perp}$.

To handle the intrinsic space-time anisotropy, 
it is convenient to introduce two
tensors 
\begin{eqnarray}
A_{\mu\nu}&=&\left(\delta_{\mu 
0}-\frac{k_{\mu}\omegan}{k^2}\right)\frac{k^2}{\bk^2}
\left(\delta_{0\nu}-\frac{\omegan k_{\nu}}{k^2}\right)~,\nonumber \\ 
B_{\mu\nu}&=&\delta_{\mu i}\left(\delta_{ij}-\frac{k_i 
k_j}{\bk^2}\right)\delta_{j\nu}~,
\end{eqnarray}
and rewrite the gauge field action as
\begin{equation}
{\cal L}_0[a_{\mu}]=\frac{1}{2}\Pi^0_A a_{\mu}A_{\mu\nu}a_{\nu} + 
\frac{1}{2}\Pi^0_B a_{\mu}B_{\mu\nu}a_{\nu}~~.
\end{equation}
In the above equations $k_{\mu}=(\omegan,\bk)$ i.e. 
$k^2=\omegan^2+\bk^2$.
It is straightforward to show that 
\begin{eqnarray}
\Pi^0_A&=&\!\! K_{\tau}f_{\tau}\left(\frac{T}{\bk},\frac{T}{\omegan},
KT,K_{\tau}T\right)
(\bk^2+\omegan^2),\nonumber \\
\Pi^0_B&=&\!\! 
K_{\tau}f_{\tau}\!\!\left(\frac{T}{\bk},\frac{T}{\omegan},
KT,K_{\tau}T\right) 
\omegan^2 + \nonumber \\
&+&  K f_{\perp}\!\!\!\left(\frac{T}{\bk},\frac{T}{\omegan},
KT,K_{\tau}T\right) \bk^2~.
\end{eqnarray}

The gauge field $a_{\mu}$ couples minimally 
to $N$ Dirac spinors  representing
nodal BCS quasiparticles \cite{ftqed} ($N=2$ for a 
single CuO$_2$ layer). 
Consequently, the resulting Lagrangian reads
\begin{equation}
{\cal L}=\psib \left(i\gamma_{\mu}\partial_{\mu}+\gamma_{\mu}a_{\mu} 
\right)\psi 
+ {\cal L}_0[a_{\mu}]~~, 
\end{equation}  
where the summation over $N$ fermion flavors is understood.
The integration over Berry gauge field
$a_\mu$ reproduces the interaction among quasiparticles
arising from the topological frustration referred to earlier. 

{\em Specific heat and QED$_3$ scaling of thermodynamics:}\\
The only lengthscales that appear in the 
thermodynamics are the thermal length $\sim v_F/T$,
and the superconducting correlation 
lengths $K$, $K_{\tau}$. At $T=0$, the
two correlation lengths $K$, $K_{\tau}$ enter only as effective short 
distance cutoffs of the theory since the
electronic action is controlled by the IR fixed point of QED$_3$.
These observations allow us to write down 
the general scaling form for the doping-dependent free energy
of the pseudogap state
\begin{equation}
{\cal F}(T;x)=-\frac{T^{3}}{v_F^2} \Phi_N\bigl ( K_{\tau}(x) T,  K(x) T\bigr ) +
{\cal E}_0(x)~,
\label{freeenergy}
\end{equation}
where $\Phi_N (x,y)$ is the thermodynamic scaling function and
${\cal E}_0(x)$ is the reference ground state energy. Note that
we still maintain $v_F =v_\Delta$ for simplicity and will
discuss Dirac cone anisotropy shortly.
In the above scaling form $K_\tau$ is directly related to the $T\to 0$
{\em finite} superconducting correlation length of the pseudogap
state, $ \xi_{\rm sc} (x)$ \cite{ftqed}. The ratio $ K_\tau / K$ 
describes
the anisotropy between time-like and space-like vortex fluctuations
and is also a function of doping $x$. The scaling expressions for
other thermodynamic functions can be derived from (\ref{freeenergy})
by taking appropriate temperature derivatives.

We are interested in the low temperature regime ($T\ll T^*$) in which 
the 
thermal length $v_F/T$ is much 
longer than $ K_{\tau}$ and $K$ ($v_F/T\gg\xi_{\rm sc}(x)$) or, 
equivalently,
in the limit $\Phi_{N}(x\rightarrow 0,y\rightarrow 0)$.
This is just the limit in which 
the free energy ${\cal F}$ (\ref{freeenergy}) approaches the free 
energy
of the finite temperature QED$_3$ and the precise form of 
(\ref{L0}) becomes unimportant as long as $f_{\tau,\perp}$ remain 
finite
as $T\to 0$. Consequently, $\Phi_N(x,y)$ should be regular at $x=y=0$ 
and
takes on the value ${\cal C}_{\rm QED}$ 
which is a universal positive numerical 
constant within QED$_3$ \cite{footii,footii'}   
(see Ref. \cite{kimleewen,wen} for discussion of related issues). 
Therefore, we finally find that, at  $T\ll T^*$,
${\cal F}=-{\cal C}_{\rm QED}  T^3/v_F^2$, 
${\cal S} = 3{\cal C}_{\rm QED}T^2/v_F^2$  and $c_v =6{\cal C}_{\rm 
QED}T^2/v_F^2$.
 
	The physical origin of the above form for 
the low temperature scaling
can be illustrated by the
expression for entropy consisting of the free fermion contribution
and the part generated solely through gauge field (vortex
fluctuation) mediated interactions \cite{footii'}:
\begin{equation}
{\cal S} = N S^0_{f}+S_{int}=N {\mathcal C}_{f} T^2 
-\frac{\partial}{\partial T}\Delta F,
\label{entropy}
\end{equation}
where, for $N=2$, ${\mathcal C}_{f}=9\zeta(3)/8\pi v_F v_{\Delta}$ and
\begin{multline}
\label{fshift}
\Delta F\!\!=\!\!\sum_{X=A,B}\int \frac{d^2 \bk}{(2\pi)^2}
\int^{\infty}_{-\infty}\!\!\!\!~\frac{d\omega}{2\pi }
\coth(\frac{\omega}{2T}) \times \\ 
\tan^{-1}
\left(
\frac{{\mathcal Im}\Pi^{F ret}_X}{{\mathcal Re}\Pi^{F ret}_X+{\mathcal 
Re}\Pi^{0 ret}_X}
\right).
\end{multline}
Note that we have traded the sum over Matsubara
frequencies $\{\omega_n\}$ for an integral 
over real frequency $\omega$.\cite{tsvelik}

The retarded fermion polarization function $\Pi^{F ret}_X$ can be 
written as
$N T {\cal P}^F_{X}(\ak/T,\omega/T)$ while the bare gauge field
polarization $\Pi^{0 ret}_X$ has the form 
$ KT^2 {\cal P}^0_{X}(\ak/T,\omega/T,K T,K_{\tau}T)$.
By rescaling momenta and frequencies appearing in Eq. (\ref{fshift}),
we see that, in the low temperature limit $KT\ll 1$ ($T\ll T^*$
or $v_F/T\gg\xi_{\rm sc}(x)$), 
the bare gauge field polarization alters {\em only the 
higher order corrections} to the leading
$S\sim T^2$ scaling of QED$_3$ which itself is determined exclusively 
by the fermion polarization. Equivalently, the finite temperature
modifications (\ref{L0}) to the original QED$_3$
theory of nodal quasiparticles interacting through vortex-antivortex
fluctuations are in effect corrections to the UV cutoff of the pure 
2D quantum electrodynamics. However, the
leading order scaling of the thermodynamics of {\em pure} QED$_3$
is {\em dominated entirely}
by the IR fixed point, and is consequently {\em independent} of the UV 
cutoff.

{\em Spin susceptibility:}\\
The ``topological'' fermion spinors $\psi$ \cite{ftqed}
allow us to express the physical spin density 
$\psi^{\dagger}_{\uparrow}\psi_{\uparrow}-\psi^{\dagger}_{\downarrow}\psi_{\downarrow}$
as a Dirac fermion density $\bar{\psi}\gamma_0 \psi$. 
Consequently, the theory connects correlations among
topologically frustrated BdG spinons to 
charge fluctuations in QED$_3$. Such correlations suppress
spin response in phase-disordered underdoped cuprates and at $T=0$
one finds $\chi\sim q^2$. Of course, within {\em pure} QED$_3$ the
charge fluctuations would still vanish even at finite $T$ since that
system is always incompressible. Within the QED$_3$ theory of cuprates,
however, we must now use the ``non-relativistic''
modified finite $T$ bare Lagrangian (\ref{L0}).
As shown bellow, this translates into a {\em finite} spin susceptibility
whose precise value reflects the static limit of 
of the scaling function $f_{\tau}$. This is in contrast to the
thermodynamics where the leading low temperature behavior was determined
by the {\em pure} ``relativistically invariant'' QED$_3$ and is
insensitive to the specific structure of $f_{\tau,\perp}$. 

To compute the spin-spin correlation 
function $\langle S_z(-k) S_z(k)\rangle$ we 
introduce an auxiliary source $J_{\mu}(k)$ and 
couple it to fermion three-current. Thus
\begin{equation}
{\cal L}[\psib,\psi,a_{\mu},J_{\mu}]=\psib 
\left(i\gamma_{\mu}\partial_{\mu}+\gamma_{\mu}(a_{\mu}+J_{\mu}) \right)\psi 
+ {\cal L}_0[a_{\mu}] 
\end{equation} 
and since it is the $z$-component of the 
spin that couples to the gauge field we have
\begin{equation}
\langle S_z(-k) S_z(k)\rangle=\frac{1}{Z[J_\mu]}\frac{\delta}{\delta 
J_0(-k)}\frac{\delta}{\delta J_0(k)} Z[J_{\mu}] \Big|_{J_{\mu}=0},
\end{equation}
Z being the quantum partition function.
Now we set $a'_{\mu}=a_{\mu}+J_{\mu}$ and 
integrate out both the fermions and the
gauge field $a'_{\mu}$. The correlations 
between $a'_{\mu}$ fields are described by the 
polarization matrix which to the order $1/N$ can be written in the 
form
$\Pi_{\mu\nu}=(\Pi_A^0 + \Pi^F_A)A_{\mu\nu}+ (\Pi_B^0 + 
\Pi^F_B)B_{\mu\nu}$.
The resulting spin correlation function is then readily found to be
\begin{equation}
\langle S_z(-k) S_z(k)\rangle=\frac{\Pi^F_A \Pi^0_A}{\Pi^F_A +\Pi^0_A}
\frac{\bk^2}{\bk^2+\omega_n^2},
\label{chi}
\end{equation} 
where $\Pi^F_A$ denotes the fermion current polarization function. 
Due to the scale invariance of the (massless) topological
fermion action, the time component of the 
retarded polarization function has the scaling form  
\begin{equation}
\Pi^{F\; ret}_A(\ak,\omega,T)=N T 
\;P^F_A\left(\frac{\ak}{T},\frac{\omega}{T}\right),
\end{equation}
where $P^F_A(x,y)$ is a universal function of its arguments and $N$ is 
the 
number of the four component Dirac fermion species ($N=2$ for a single
CuO$_2$ layer). 
In the static limit, $\omega \rightarrow 0$,
\begin{equation}
\lim\limits_{y \rightarrow 0} {\mathfrak Re}\; 
P^F_A(x,y)=\frac{2\ln2}{\pi}+\frac{x^2}{24\pi} + 
{\cal O}(x^3); \;\;\;x\ll1
\end{equation}
and
\begin{equation}
\lim\limits_{y \rightarrow 0}{\mathfrak Re}\; 
P^F_A(x,y)=\frac{x}{8}+\frac{6\zeta(3)}{\pi x^2}+ 
{\cal O}(x^{-3});\;\;\;x\gg1
\end{equation}
while
\begin{equation}
\lim\limits_{y \rightarrow 0} {\mathfrak Im}\; 
P^F_A(x,y)=\frac{2\ln2}{\pi}\frac{y}{x}+ 
{\cal O}\left(\frac{y^2}{x^2}\right); \;\;\;x\ll1~.
\end{equation}

Furthermore, in order to complete the computation of spin
susceptibility, we observe that
in the limit $\omega =0$ and $\bk \rightarrow 0$ we have
\begin{equation}
f_{\tau}\left(\frac{T}{\ak},\infty, 
KT,K_{\tau}T\right)\rightarrow 1+ c \frac{T^2}{\bk^2}
\label{ftau}
\end{equation}
where $c$ is a pure number \cite{footi}. The expression (\ref{ftau})
for the static limit
of $f_\tau$ is rather general and can be understood on 
physical grounds: in the 
pseudogap state at $T\to 0$ we can think of the ``normal'' system as a dual 
superfluid of
relativistic ``vortex bosons'' \cite{ftqed}. These bosons are coupled
to a massless dual gauge field which describes the original long
range vortex-vortex interactions. Once $T$ is finite, this dual
superfluid immediately losses its off-diagonal order since the
screening by the dual gauge field allows for creation of a finite
density of thermally activated dual vortices, {\em i.e.} ``vortices'' in
the dual superfluid of original vortices. Thus, there is
no finite $T$ transition in the dual superfluid that would be 
analogous to the Kosterlitz-Thouless transition in original superconductor.
The above form for $f_\tau$
(\ref{ftau}) is just the mathematical expression of this fact.
 
Combining all of these results together we finally obtain
the spin susceptibility:
\begin{equation}
\chi(\omega=0,\ak\rightarrow 0)= 
\frac{2N\ln2}{\pi}\frac{ K_{\tau}T^2}{\frac{2N\ln2}{\pi c}+ 
K_{\tau}T}~~.
\label{susceptibility}
\end{equation}
Consequently, 
Wilson ratio $R=\chi T/c_v$ vanishes as $T\to 0$ \cite{footi}
implying the non-Fermi liquid nature  
of the pseudogap state in cuprates within the theory of Ref. \cite{ftqed}.

{\em Corrections to scaling:}\\
There are several sources of corrections to the QED$_3$ scaling, and it is 
not practical to address them all within the confines of this Letter.
Instead, we choose to concentrate here 
on those particular corrections to scaling which arise from 
the Dirac cone anisotropy. 
Such corrections are most likely to
be important in real materials where the values of velocities
$v_F$ and $v_\Delta$ as functions of underdoping are now becoming
known with increasing precision \cite{taillefer}. 
It was shown in Ref. \cite{vtf} that the Dirac cone anisotropy 
$\alpha_D=v_F/v_{\Delta}$ 
scales to unity at the QED$_3$ IR fixed point. Alternatively, when 
defined via
$\alpha_D \equiv 1+\delta$, $\delta$ has a small, but finite 
scaling dimension $\eta_{\delta}>0$. 
To the leading order in $1/N$, $\eta_{\delta}=32/(5\pi^2N)$ 
\cite{vtf}.
For simplicity, we assume here that the bare action for the gauge field is 
$e^{-2} F_{\mu\nu}^2$ ($K=e^{-2}$), where 
$F_{\mu\nu}=\partial_\mu a_\nu -\partial_\nu a_\mu$. 
Then the $T-$dependent part of the free energy scales as
\begin{equation}
{\cal F}=-\frac{T^3}{v_F^2} 
\Phi_N\left(\frac{T}{e^2},\frac{v_F}{v_\Delta}\right)=
-\frac{T^3}{v_F^2} \Phi_N\left(\frac{T}{e^2},1+\delta(T) \right),
\end{equation}
and in the limit of $T\rightarrow 0$ we have
\begin{equation}
{\cal F}=-\frac{T^3}{v_F^2} 
\Phi_N\left(0,1\right)-\frac{T^{3+\eta_{\delta}}}{v_F^2}\delta_0
\Phi_N'\left(0,1\right)~~,
\end{equation}
where $\Phi_N$ and $\Phi_N'$ are pure numbers and $\delta_0=\alpha_D-1$.
Thus, although the leading order scaling in $T$ is analytic, just
like in the pure finite-$T$ QED$_3$, the fact 
that the next-to-leading order correction is not reflects the strongly
interacting nature of the QED$_3$ IR fixed point. 

In summary, we have shown that the leading order low
temperature scaling of the thermodynamic
functions in the pseudogap state is that of a {\em pure} finite-$T$ QED$_3$, 
since the finite temperature modifications (\ref{L0}) to the theory
of Ref. \cite{ftqed} affect only corrections to scaling. 
As a result, the specific heat goes as  $c_v \sim T^2$.
Furthermore, we argued that at low temperatures 
the spin susceptibility is suppressed,
$\chi \alt T^2$, due to correlations among BdG spinons mediated
by an emergent gauge field $a_\mu$. 
This implies a vanishing Wilson ratio and
breakdown of the Fermi liquid
behavior within the pseudogap state. 
Finally, the corrections to scaling arising from Dirac 
cone anisotropy were argued to be non-analytic, representing yet another
manifestation of the strongly interacting
character of the QED$_3$ brand of quantum criticality.

This work was supported in part by the NSF grant DMR00-94981.


\begin{references}
\bibitem{plasmon} In the absence of long range Coulomb interactions, there
is an additional contribution to the low $T$ thermodynamics from the
Bogoliubov soft phase mode. However, this mode is lifted to plasma frequency
once Coulomb interactions are accounted for and therefore it will 
not change the leading low-$T$ thermodynamics.
\bibitem{footiv} One finds that the pseudogap state is
a charge dielectric by retracing the reasoning that leads to Eq. 
(\ref{susceptibility}) but for the Doppler gauge field $v_\mu$.\cite{ftqed}
\bibitem{footii} ${\cal C}_{\rm QED}$ depends only on $N$,
the number of Dirac fermion species.
We are unable to determine ${\cal C}_{\rm QED}$
analytically and will report on its numerical computation elsewhere.
\bibitem{footii'} The original derivation of 
QED$_3$ theory \cite{ftqed}
mandates that the ``free photon'' contribution be absent from
${\cal C}_{\rm QED}$ (and $\Delta F$ (\ref{fshift}))
which contains only the parts contributed by 
free fermions and their interactions mediated by $a_\mu$.
\bibitem{footi} Strictly speaking $c$ is a function of $KT$ and
$K_\tau T$ and we cannot exclude its vanishing 
in the $KT\to 0$ limit, depending on
a specific model for the dual superfluid. In this sense our
expression for spin susceptibility (\ref{susceptibility})
should be regarded as only an upper limit. Note that the thermodynamics
remain unaffected by the specific form of  $f_{\perp,\tau}$
as long as the $T\to 0$ behavior defined below Eq. (\ref{L0})
remains in force. Therefore, the vanishing Wilson ratio for  
$T\to 0$ is a rather general result of our theory.
\end{references}
\end{document}